\documentstyle[12pt]{article}

\def\s{\sigma}

\def\unlock{\catcode`@=11 }
\def\lock{\catcode`@=12 }
\unlock
\def\@@eqncr{\let\@tempa\relax\global\advance\@eqcnt by \@ne
    \ifcase\@eqcnt \def\@tempa{& & & &}\or \def\@tempa{& & &}\or
     \def\@tempa{& &}\or \def\@tempa{&}\else\fi
     \@tempa \if@eqnsw\@eqnnum\stepcounter{equation}\fi
     \if@defeqnsw\global\@eqnswtrue\else\global\@eqnswfalse\fi
     \global\@eqcnt\z@\cr}

\def\@eqnacr{{\ifnum0=`}\fi\@ifstar{\@yeqnacr}{\@yeqnacr}}
\def\@yeqnacr{\@ifnextchar [{\@xeqnacr}{\@xeqnacr[\z@]}}
\def\@xeqnacr[#1]{\ifnum0=`{\fi}\cr \noalign{\vskip\jot\vskip #1\relax}}
\def\eqalign{\null\,\vcenter\bgroup\openup1\jot \m@th \let\\=\@eqnacr
\ialign\bgroup\strut
\hfil$\displaystyle{##}$&$\displaystyle{{}##}$\hfil\crcr}
\def\endeqalign{\crcr\egroup\egroup\,}

\def\cases{\left\{\,\vcenter\bgroup\normalbaselines\m@th \let\\=\@eqnacr
    \ialign\bgroup$##\hfil$&\quad##\hfil\crcr}
\def\endcases{\crcr\egroup\egroup\right.}
\lock

\def\nast{\stackrel{*}{N}\hspace{-3.5pt}\vphantom{N}}
\begin{document}
\unitlength=1mm

\vspace{20mm}

\title{
Variational principle and a perturbative solution of non-linear string
equations in curved space }
\author{\sl  S.N. Roshchupkin \\
  Simferopol State University, ~333036, Simferopol, Ukraine\\
and \\
\sl  A.A. Zheltukhin\thanks{E-mail: zheltukhin@kipt.kharkov.ua}\\
 Kharkov Institute of Physics and Technology\\
 310108, Kharkov, Ukraine}
 \date{}
 \maketitle

\begin{abstract}
String dynamics in a curved space-time is studied
on the basis of an action functional including a small parameter of
rescaled tension $\varepsilon=\gamma/\alpha^{\prime}$ , where $\gamma$ is
a metric parametrizing constant. A rescaled slow worldsheet time
$T=\varepsilon\tau$ is introduced, and general covariant  non-linear
string equation are derived.

It is shown that in the first order of an
$\varepsilon-$expansion these equations are reduced to the known
equation for geodesic derivation but complemented by a string oscillatory
term. These equations are solved for the de Sitter and Friedmann -Robertson-
Walker spaces. The primary string constraints are found to be split into a
chain of perturbative constraints and their conservation and consistency are
proved. It is established that in the proposed realization of the
perturbative approach the string dynamics in the de Sitter space is
stable for a large Hubble constant $H \,\, (\alpha^{\prime}H^{2}\gg1)$.
\end{abstract}

\bigskip
  \thispagestyle{empty} \section{Introduction}

In recent years much attention has been paid to studying the role of
strings in cosmology [1-3]. Investigation of this problem is complicated
by nonlinear character of string equations solvable for special types of
metrics. Therefore in [3-6] an approach to studying
approximate solutions of string equations using a perturbative expansion
was initiated. This approach is based on the idea of an expansion of string
solutions around the geodesic line of the string mass center described by a
mass parameter $m $. A great deal of work has been done in this direction,
and a considerable class of perturbative string equation solutions was
found for different cosmological spaces [3-13 and Refs. there].

An extensive application of this perturbative approach necessitates its
further investigations. In particular, the nature of a small perturbative
parameter and the procedure of its bringing into the string equations and
constraints are important points for study.  Moreover, it is appropriate
to find a mechanism for fixing arbitrariness in the choice of the
phenomenological mass parameter $m $ , to define a relevant scale for
measuring the worldsheet parameter $\tau$ and $\sigma$ which are the
arguments of perturbative functions. In principle a well defined mass
parameter attributed to the center mass trajectory may be absent. For
example, in contrast to the case of the Minkowski space-time, where a
particle is characterized by a fixed mass and spin, in a curved space-time
a particle has some fixed eigenvalues of other Casimir operators.
These operators are built of the generators of a symmetry group of the
curved space-time and may be a complicated combination of the momentum
and spin operators. To investigate the above mentioned problems seems
to be important for the classification of cosmological spaces where a
perturbative string dynamics is selfconsistent.

While studying this matter, a new representation for the string
action including kinetic and potential terms of the string lagrangian as
independent additive terms was  considered in [14] \footnote{This
representation is a natural generalization of the representation [15]
to the case of an arbitrary curved space-time.}. This representation
comprises a rescaled string tension $\gamma/\alpha^{\prime}$ as a
small dimensionless parameter -- a world sheet "cosmological term". The
constant $\gamma $ in the rescaled tension $ \gamma/\alpha^{\prime}$ with
the dimension $L^2 \, (\hbar=c=1)$ is a constant parameterizing the
metric of a curved space-time. For example, for the de Sitter space
$\gamma=H^{-2}$, where H is the Hubble parameter.  Using this
representation for the Nambu-Goto string action the perturbative string
equations were derived. The perturbative string equtions [14], were shown
to be transformed into the perturbative
equations [ 4,5] after rescaling the worldsheet parameter $\sigma$ (or
$\tau$) and fixing the phenomenological mass parameter by the value $m=0$.
These results point out to the existence of different realizations of the
considered perturbative approach. So it becomes important to establish the
regions  for applicability of the different  realizations  and to
understand  the physical effects connected with them. In particular, it
may occur that the perturbative string dynamics critically depends on the
value of the phenomenological mass parameter $m $ for some type of the
curved space-time.  The de Sitter space just belongs to this case.
Actually, as shown in [5], the perturbative string frequency modes in the
de Sitter space are defined as $\omega_{n}=\sqrt{n^{2}-(\alpha'Hm)^{2}}$
and become imaginary for large values of the Hubble constant H. This
results in instabilities of the string dynamics in the realization of the
 perturbative approach considered in [4,5].  It follows from the above
formula for $\omega_{n}$ that these instabilities must disappear, if the
phenomenological parameter $m$ acquires zero value. This value is in exact
accordance with the restriction of the perturbative scheme realized in
[14].  Therefore it seems important to present a rigorous verification of
the absence of instabilities in the realization of the perturbative
approach proposed in [14], as well as to develop and substantiate this
perturbative scheme itself.

Note also that in [16] a perturbative approach to strings using null
string as zero approximation was considered. As a result the perturbative
equations [16] did not include any oscillatory terms in the first
and second approximations. In view of the fact that the introduction of any
arbitrary small tension should excite an oscillatory regime in the pattern
of the string evolution, it becomes obvious that the realization [16] of
the perturbative approach needs additional groundings.

 A novelty of the present paper is the introduction of a rescaled slow
 worldsheet parameter $T=\varepsilon\tau$, where
 $\varepsilon=\gamma/\alpha^{\prime}$ is a small dimensionless parameter
 presenting the rescaled string tension.  The transition to the scale $T$
 shows the degree of roughening of the string dynamics in the considered
 perturbative scheme. Using general covariant formulation of the
 perturbative equations and constraints we show that the string equations
 in the first approximation acquire the form of the geodesic deviation
 equations [22] complemented by an additional oscillatory term .

   It is proved that primary non-perturbative constraints are
split into a system of constraints for the perturbative functions. We
 show that the general covariantization procedure provides an essential
 simplification of these perturbative constraints. The proof of the
 consistency of these perturbative constraints and their conservation is
 presented. This proof becomes very simple in the proposed general
 covariant formulation. Further we find that the constraints of the
 first approximation functions are reduced to the condition of their
 orthogonality to the geodesic world trajectories of the zero
 approximation.  We establish that the constraints in question may be
 considered as the initial data of the perturbative equations.

 Considered  is the application of the perturbative approach for a wide class
  of the Friedmann-Robertson-Walker universes. It
 is shown that their linearized equations of the first approximation have
 the form of the modified Bessel equations. Their exact solutions
 are found.

\thispagestyle{empty} \section {Rescaled tension as a perturbation
\protect\\ in the Nambu-Goto action}

As shown in [14,15], the Nambu-Goto string action in the curved space can
be presented in the form
 \begin{equation}\label{1}
 S=S_0+S_1 = \int d\tau d\sigma
\left[ {det (\partial_\mu x^M G_{MN}(x) \partial_\nu x^N)\over
E(\tau,\sigma)} - {1\over(\alpha^\prime)^2} E (\tau ,\sigma)\right]~~,
\end{equation}
 where $E$ is an auxiliary world-sheet density. The motion equation for $E$
 produced by $S$ (1) is
\begin{equation}\label{2}
E= \alpha^\prime \sqrt{-\mbox{det } g_{\mu\nu}} ,
\end{equation}
\begin{equation}\label{3}
g_{\mu\nu} = \partial_\mu x^M G_{MN}(x)\partial_\nu x^N ,
\end{equation}
The substitution of $E$  (2) into the functional $S$  (1) transforms  the
latter into the Nambu-Goto representation
  \begin{equation}\label{4}
S=-\frac{2}{\alpha^{\prime}}\int d\tau d\sigma \, \sqrt{-\mbox{det }\left(
\partial_\mu x^M G_{MN}(x)\partial_\nu x^N \right)}
\end{equation}
 Thus, the representations (1) and (4) for the string action $S$ (1) are
classically equivalent. Unlike the  representation (4), the
representation (1) includes the string tension  parameter
$1/\alpha^{\prime}$ as a constant at an additive world-sheet
"cosmological" term playing the role of potential energy. Respectively,
this term may be considered as a perturbative addition for the case of a
week tension. But what are the measure units in terms of which the string
tension is a small value?

To answer this question we  are to consider one of dimensional parameter
$\gamma$ or some combination of the parameters defining the metric of the
curved space where the string moves. Without loss of generality put that
$\gamma$ has the dimension $ L^2 (\hbar=c=1) $. Then the value of the
dimensionless combination \begin{equation}\label{5}
\varepsilon=\gamma/\alpha^{\prime}
\end{equation}
can be considered as a parameter characterizing  the power of string
tension. When  $\varepsilon\ll 1$, or equivalently,
\begin{equation}\label{6}
1/\alpha^{\prime}\ll \gamma^{-1} ,
\end{equation}
the tension  $1/\alpha^{\prime}$ should be considered as a weak one.
For example, in the de Sitter space the Hubble parameter $H$ plays the
role of $\gamma^{-1/2}$, and we consider  tension as a weak one when
\begin{equation}\label{7}
1/\alpha^{\prime}\ll H^{2}
\end{equation}

Of course, for the cases of more complicated background  including
additional fields such as members of supergravity multiplet, we get
wider  possibilities for the choice of a perturbative parameter.

 A natural condition for appearance  of $\varepsilon$ (5) in the
representation (1) is the agreement that the string world coordinates
$x^{M} $ are measured in terms of the metric parameter $\gamma$. Actually,
if we choose dimensionless coordinates ${\tilde x}\vphantom{x}^{M}$ and
the Lagrange multiplier $\tilde{E}$
\begin{equation}\label{8}
x^{M}=\gamma^{1/2}{\tilde x}\vphantom{x}^{M} \quad,\quad
E=\gamma^{2}\tilde{E}
\end{equation}
the action S  $(1)$ is presented in the form
\begin{equation}\label{9}
S= \int d\tau d\sigma \left[ \frac{\mbox{det } (\partial_\mu {\tilde x}^M
G_{MN}(\tilde x)
\partial_\nu
{\tilde x}^N)}{\tilde E} - \left( \frac{\gamma}{\alpha^\prime}\right)^2 \tilde{E}
\right] \end{equation}
containing the dimensionless parameter  $\varepsilon$ .  In the case, when
$x^{M} $ are measured by the constant $ \alpha^{\prime}$, i.e.
\begin{equation}\label{10}
x^{M}=\sqrt{\alpha^{\prime}}{\bar x}\vphantom{x}^{M} \quad,\quad
E={\alpha^{\prime}}^{2}{\bar E}
\end{equation}
the "cosmological term" in the representation of the action (1) loses
the role of a perturbation term.  If we prefer to work in terms of the
original world coordinates $x^M$, then the condition for the measurement
of $x^M$ in the units of the constant $\gamma$ is manifested by the choice
of a worldsheet gauge fixing in the form [14] \begin{equation}\label{11}
E=-\gamma(\acute{x}\vphantom{x}^{M}G_{MN}\acute{x}\vphantom{x}^{N})
\end{equation}
In the gauge (11) complemented by the orthonormality condition
\begin{equation}\label{12}
\left( \dot{x}\vphantom{x}^{M}G_{MN}\acute{x}\vphantom{x}^{N} \right)=0
\end{equation}
the variational Euler-Lagrange motion equations generated by $S$ (1)
acquire the form [14]
\begin{equation}\label{13}
{\mathaccent "7F x}^M -
\left({\gamma \over \alpha^\prime }\right)^2 \stackrel{\scriptstyle
\prime\prime}{x} \hspace{-3.5pt}\vphantom{x}^{M}
+\Gamma^M_{PQ}(x)\left[\dot{x}^P\dot{x}^Q-\left({\gamma \over \alpha^\prime }
\right)^2 \acute{x}\vphantom{x}^{P}\acute{x}\vphantom{x}^{Q} \right]=0
\end{equation}
and contain the dimensionless parameter $\varepsilon$ (5). This parameter
appears in another string constraint
\begin{equation}\label{14}
\left( \dot{x}\vphantom{x}^{M}G_{MN}\dot{x}\vphantom{x}^{N} \right) +
\left(\frac{\gamma}{\alpha^{\prime}} \right)^{2}
\left( \acute{x}\vphantom{x}^{M}G_{MN}\acute{x}\vphantom{x}^{N} \right)=0
\end{equation}
which is additional to (12) and follows from Eqs.(2) and (11). Provided
that ${\gamma \over \alpha^\prime}\ll 1$ Eqs.(14) can be considered as
nonlinear equations with the small parameter $\varepsilon$ (5). Then we
should seek for a solution of (13) and for the constraints (12) and
(14) in the form of a series expansion in terms of $\varepsilon$ (5).

Physically the case  $\varepsilon\ll 1$ corresponds to a strong
gravitational field  or, equivalently, to a large scalar space-time
curvature $R^{\ M}_{M} $ measured in terms of the tension
$1/\alpha^{\prime}$. This is
evident for the case  of the de Sitter spaces where the condition (6) is
equivalently presented in the form \begin{equation}\label{15}
1/\alpha^{\prime} \ll R^{\ M}_{M} ,
\end{equation}
which shows that the elastic force  of the string is less than the gravity
force.  In the limit of zero tension $\varepsilon=0$
$(\alpha^{\prime}\rightarrow \infty) $ the action (1), constraints (12-14)
and Eqs.(13) transform into the relations characterizing a tensionless
string or a massless particle.  The tensionless string moves
translatingly along the the geodesic lines of the considered space-time
without any oscillations. In the case of $\varepsilon\ll 1$ a very small
elastic force described by the terms  with the $\sigma-$derivatives in
(13-14) appears  in addition to the external gravity force. Then the small
string oscillations  appear  and each point of string gets  an additional
shift. But the amplitudes of these oscillator shifts are smaller than the
paths caused by the translating movements. Thus these oscillations can be
considered  as small perturbations of the  translating movements of the
string points.

The perturbative oscillations are characterized by small frequencies and,
subsequently, by large periods. A  characteristic  time scale of the
oscillator periods is proportional to $ 1/\varepsilon$. This observation
follows from the string equations (13)  where the transition to a
rescaled worldsheet proper time  $T$ \begin{equation}\label{16}
T=\varepsilon\tau \quad,\quad
\frac{\partial}{\partial\tau}=\varepsilon
\frac{\partial}{\partial T} \quad,\quad
\frac{\partial^{2}}{\partial\tau^{2}}=\varepsilon^{2}
\frac{\partial^{2}}{\partial T^{2}}
\end{equation}
is performed. Such a transition transforms Eqs.(13) and the constraints
(12,14) into the standard form
\begin{equation}\label{17}
x^{M}_{,TT}-\stackrel{\scriptstyle\prime\prime}{x}
\hspace{-3.5pt}\vphantom{x}^{M}+
\Gamma^{M}_{PQ}(x) \left[ x^{P}_{,T}x^{Q}_{,T}-\acute{x}\vphantom{x}^{P}
\acute{x}\vphantom{x}^{Q} \right] =0
\end{equation}
\begin{equation}\label{18}
\left( x_{,T}^{M}\acute{x}_{M} \right) =0
\end{equation}
\begin{equation}\label{19}
x^{M}_{,TT}x_{M,TT}+\acute{x}\vphantom{x}^{M}\acute{x}_{M}=0 ,
\end{equation}
where $ x_{,T}^{M}\equiv\partial_{T}{x}^{M}$ and
$x_{M,T}\equiv G_{MN}\partial_{T}{x}^{N}$. The transition to the slow
worldsheet time $T $ (16) means  an enlargement of the original world
sheet time $\tau$ by $1/\varepsilon$ times. The choice of such large units
for the worldsheet time leads to an essential roughening of the string
motion pattern due to which an information on the microscopic string
dynamics is lost. On the slow  scale  $T $ the string oscillations can be
observed owing to  a sufficient observation time. But the rescaling of the
worldsheet time does not result in an increase of oscillation
amplitudes in contrast to the translating  movement length. Thus, after
the exclusion of the small parameter $\varepsilon$ from the string
equations and the constraints (17-19) the ratio of the oscillation
amplitudes to the translation displacement plays the role of a small
parameter. This allows to seek  for the string equations
solution on the scale $T $ in the form of a superposition of its large
translating and small oscillation displacements
\begin{equation}\label{20}
x^M =
\varphi^M(T)+\varepsilon\psi^M(T,\sigma)+\varepsilon^2\chi^M(T,\sigma)
+ ...
\end{equation}

The zero approximation functions $\varphi^M$ in the asymptotic series
expansion (20) do not depend on the parameter $\sigma$ which enumerates
different points of the string.  Such a choice is explained by  an
assumption that differences  in the displacements of the string points and
the oscillation amplitudes have the same order of smallness.  While
comparing Eqs.(17) with the correspond ones in [4] we conclude that
the latter transform into (17) after  a formal change $ \tau \rightarrow
T$.  This observation means that from the view point of the variational
principle for the string  action (1), the perturbative  expansion [14]
works starting from  $\tau\geq1/\varepsilon$.

The perturbative equations  and constraints  following from the
exact  ones (17-19) have been  derived in [14] in terms of
the worldsheet variables $(\tau, \xi)$, where $ \xi=\sigma/\varepsilon $.
The variables $(\tau, \xi) $ are connected with the variables
$(T, \sigma)$ used  here by the dilaton transformation defined by the
parameter $\varepsilon$
\begin{equation}\label{21} T=\varepsilon\tau
\quad,\quad \sigma=\varepsilon\xi
\end{equation}
Usage of the  variables
$(\tau, \xi) $ in the perturbative description is not very
convenient due  to the appearance  of $\varepsilon$ in the boundary
conditions  for closed string discussed here
\begin{equation}\label{22}
x^M(T ,\s =0) = x^M(T ,\s=2\pi)
 \end{equation}
  This inconvenience
disappears when the variables $(T, \sigma)$  (16-19) are introduced. Take
into account the fact that the transformation (21) belongs to the two
dimensional conformal group  which is  a local symmetry  of the string
equations and constraints. At the same time  we find that the
perturbative equations and constraints  generated by Eqs.(17-19) are
obtained from the correspond ones in [14] after the simple change $(\tau,
\xi) \rightarrow (T,\sigma)$.

 For the zero approximation functions $\varphi^{M}(T)$ we get the
equations
\begin{equation}\label{23}
\varphi^{M}_{,TT}+\Gamma^{M}_{PQ}(\varphi)\varphi^{P}_{,T}\varphi^{Q}_{,T}=0 ,
\end{equation}
$$
\left( \varphi^{M}_{,T}\varphi_{M,T}\right) \equiv
\left( \varphi^{M}_{,T}G_{MN}(\varphi)\varphi_{,T}^{N}\right) =0 \eqno(23') $$
The constraint  $(23')$ shows that
the vector $\varphi^{M}_{,T}$ is a light-like vector corresponding to
4-velocity of a massless particle  moving along the geodesic  line (23).
Thus we find that the variational principle  applied to  the action (1)
fixes the value of the mass parameter $m $ introduced  in [4,5].
Later we will see  that this fixation $m=0 $ leads to important
concequences.

The equations and constraints for the first approximation functions
$\psi^{M}(T,\sigma)$ take the form
\begin{equation}\label{24}
\Delta_{L}^{M}\psi^{L} \equiv \psi^{M}_{,TT}-
\stackrel{\scriptstyle\prime\prime}{\psi}\hspace{-3.5pt}
\vphantom{\psi}^{M}+2\left[ \Gamma^{M}_{PQ}(\varphi)\varphi^{P}_{,T}
\psi^{Q}_{,T}+
\frac{1}{2}\psi^{L}\partial_{L}\Gamma^{M}_{PQ}
\varphi^{P}_{,T}\varphi^{Q}_{,T}\right] =0,
\end{equation}
$$
\left( \varphi_{M,T}\psi^{M}_{,T}\right)+\frac{1}{2}\psi^{L}\left(
\varphi^{M}_{,T}\partial_{L} G_{MN}\varphi^{N}_{,T}\right)=0 , \eqno(24')$$
\begin{equation}\label{25}
\left(\varphi_{M,T}\acute{\psi}\vphantom{\psi}^{M}\right) =0
\end{equation}
Since the constraint $(24')$  is a constraint   for the initial
data of Eqs.(24) we may integrate $(24')$ with respect to $\sigma$ and
obtain
  $$\left( \varphi_{M,T} {\psi}^{M}\right) =C(T),$$
because $G_{ML}\varphi^{L}_{,T}$ do not depend on the variable $\sigma$.
The kinematic "constant"  $C(T)$ may be chosen equal to zero without loss of
generality. This choice means that the perturbation $\psi^{M}$ produced by
the  small string oscillations is orthogonal  to the light-like
geodesic line (23). Further  we shall use the orthogonality constraint
$$
\left( \varphi_{M,T} {\psi}^{M}\right) =0     \eqno(24'')
$$
instead of (25), and it is a new point in comparison with the results
of [14].

Finally  the equations and constraints for the functions of the
second-order approximation $\chi^{M}(T, \sigma)$ in terms of $(T, \sigma)$
worldsheet variables  get the form
\begin{equation}\label{26}
\Delta^{M}_{L}\chi^{L}+\Gamma^{M}_{PQ}\left[ \psi^{P}_{,T}\psi^{Q}_{,T}
-\acute{\psi}\vphantom{\psi}^{P}\acute{\psi}\vphantom{\psi}^{Q} \right]+
2\psi^{L}\partial_{L}\Gamma^{M}_{PQ}\varphi^{P}_{,T}\psi^{Q}_{,T}+
\frac{1}{2}\psi^{L}\psi^{K}\partial_{LK}
\Gamma^{M}_{PQ}\varphi^{P}_{,T}\varphi^{Q}_{,T}=0 ,
\end{equation}
$$
2\left( \varphi_{M,T}\acute{\chi}\vphantom{\chi}^{M}\right)
+\chi^{L}\partial_{L}G_{MN}\varphi^{M}_{,T}\varphi^{N}_{,T}+\left[\left(
\psi^{M}_{,T}\psi_{M,T}\right]+ \left( \acute{\psi}\vphantom{\psi}^{M}
\acute{\psi}_{M}\right)\right] + $$
$$
2\psi^{L}\partial_{L}G_{MN}\varphi^{M}_{,T}\psi^{N}_{,T}+
\frac{1}{2}\psi^{L}\psi^{K}\partial_{L}
G_{MN}\varphi^{M}_{,T}\varphi^{N}_{,T}=0 ,\eqno(26')$$
$$ \left(\varphi_{M,T}\acute{\chi}\vphantom{\chi}^{M}\right)+\left(
\psi_{M,T}\acute{\psi}\vphantom{\psi}^{M}\right)+
\psi^{L}\partial_{L}G_{MN}\varphi^{M}_{,T}\acute{\psi}\vphantom{\psi}^{N}=0
\eqno (26'') $$
Eqs.(23-26) should be complemented by the periodicity conditions  for
 $\psi^{M}$ and $\chi^{M}$
\begin{equation}\label{27}
\begin{array}{l}

\psi^{M}(T,\sigma=0)=\psi^{M}(T,\sigma=2\pi) , \\
\chi^{M}(T,\sigma=0)=\chi^{M}(T,\sigma=2\pi)
\end{array}
\end{equation}
Eqs.(23-26) show that the effects caused by the appearance of a
small tension  manifest themselves starting from the first
approximation (and conserve in the second one). On the scale $T$, i.e.,
when $\tau\geq1/\varepsilon$, these effects have an oscillation character
and agree with  the qualitative picture  described here  and in [14].

\section{General covariance and consistency of the perturbative
constraints and equations}

In this section we show a general covariant character of the perturbative
scheme  under discussion and  prove the selfconsistency of the
perturbative split chain of the equations and constraints (23-26).

The general covariant  differential and derivative corresponding to a
target space metric $G_{MN}(\varphi)$ are
 \begin{equation}\label{28}
 \begin{eqalign}
{\cal D}V^{M}=dV^{M}+d\varphi^{P} \Gamma^{M}_{PQ}(\varphi)V^{Q}\\
{\cal D}_{T}V^{M}=V^{M}_{,T}+\varphi^{P}_{,T} \Gamma^{M}_{PQ}(\varphi)V^{Q}
\end{eqalign}
\end{equation}
The definition (28) turns out to present  the geodesic equation
(23) in the form \begin{equation}\label{29} {\cal
D}_{T}\varphi^{M}_{,T}=0, \end{equation}

To prove the conservation of the constraint $(23')$ let us differentiate
it with respect to $\tau$ and get
\begin{equation}\label{30}
\partial_{T}\left( \varphi^{M}_{,T}G_{MN}(\varphi) \varphi^{N}_{,T}\right)=
2\left( {\cal D}_{T}\varphi^{M}_{,T}\cdot\varphi_{M,T}\right)+
\varphi^{M}_{,T}\varphi^{N}_{,T}{\cal D}_{T}G_{MN}(\varphi)
\end{equation}
The first and the second terms in (30) equal zero in view of (29),
and the the well known property of $G_{MN}(\varphi)$
\begin{equation}\label{31}
{\cal D}_{T}G_{MN}=0 ,
\end{equation}

respectively.
 The motion equations (24) for the first order perturbative  functions
 $\psi^{M}(T, \sigma)$ involve the  differential operator $\Delta_{L}^{M}$
\begin{equation}\label{32}
\Delta_{L}^{M}\equiv \delta^{M}_{L}(\partial^{2}_{T}-\partial^{2}_{\sigma})+
2\varphi^{P}_{,T}\Gamma^{M}_{PQ}(\varphi)\partial_{T}+\varphi^{P}_{,T}
\varphi^{Q}_{,T}\partial_{L}\Gamma^{M}_{PQ}
\end{equation}
Using the definition of $\Gamma^{M}_{PQ} $ [22] and their independence on
$\sigma$ we find that the general covariant representation of
 $\Delta_{L}^{M}$ has the form
\begin{equation}\label{33}
\Delta^{M}_{L}=\delta^{M}_{L}\left( {\cal D}^{2}_{T}-{\cal D}^{2}_{\sigma}
\right) -R^{M}_{\ PQL}\varphi^{P}_{,T}\varphi^{Q}_{,T},
\end{equation}
where $R^{M}_{\ PQL}$ is the Riemann-Christoffel tensor
\begin{equation}\label{34}
\frac{1}{2}R^{M}_{\ PQL}=\partial_{[Q}\Gamma^{M}_{L]P}+
\Gamma^{N}_{P[L}\Gamma^{M}_{Q]N}
\end{equation}
Note that  ${\cal D}^{2}_{\sigma}=\partial^{2}_{\sigma}$  since
$\varphi^{M}(T)$ is independent on $\sigma$. By means of these
observations  and definitions the equations and constraints $(24-24'')$
can be rewritten in the general covariant form
\begin{equation}\label{35}
\left( {\cal D}^{2}_{T}-{\cal D}^{2}_{\sigma}\right)\psi^{M} +
R^{M}_{\ PQL}\varphi^{P}_{,T}\varphi^{Q}_{,T}\psi^{L}=0
\end{equation}
$$
\left( \varphi_{M,T}{\cal D}_{T}\psi^{M}\right)=0 ,            \eqno(35')$$
$$
\left( \varphi_{M,T}\psi^{M}\right)=0 ,
\eqno(35'')$$
In the absence of the term ${\cal
 D}^{2}_{\sigma}\psi^{M}$ in Eqs.(35) the latter acquire the form of the
geodesic deviation equations [22]. The term
${\cal D}^{2}_{\sigma}\psi^{M}=\partial^{2}_{\sigma}\psi^{M}$ in (35)
describes the contribution of the string elastic forces  which push out
the string points from the geodesic lines enumerated by the parameter
$\sigma$.

To prove the conservation of the constraints $(35')$ we are to
differentiate them
\begin{equation}\label{36} \partial_{T}\left(
\varphi_{M,T}{\cal D}_{T}\psi^{M}\right)= \left( {\cal
D}_{T}\varphi_{M,T}{\cal D}_{T}\psi^{M}\right)+ \left( \varphi_{M,T}{\cal
D}^{2}_{T}\psi^{M}\right) \end{equation} The first term in (36) equals to
zero because of the Eqs.(29). After using Eqs.(35) the second term in
(36) takes the form
\begin{equation}\label{37} \left( \varphi_{M,T}{\cal
D}^{2}_{T}\psi^{M}\right)= \left(
\varphi_{M,T}\partial^{2}_{\sigma}\psi^{M}\right)- \varphi_{M,T} R^{M}_{\
PQL}\varphi^{P}_{,T}\varphi^{Q}_{,T}\psi^{L}= \partial^{2}_{\sigma}\left(
\varphi_{M,T}\psi^{M}\right) \end{equation} and also goes to zero owing
to the constraint $(35'')$.

 Similar reasoning should be used to prove the conservation of
 the constraint $(35'')$ which takes the form
 \begin{equation}\label{38}
 \partial_{T} \left( \varphi_{M,T}\psi^{M}\right)= \left(
 \varphi_{M,T}{\cal D}_{T}\psi^{M}\right)
 \end{equation}
 after differentiation and taking into account of Eqs.(29) and (31). The
right-hand side of (38)equals zero in view of the constraints $(35')$.
Thus, the constraints $(23')$ and $(35', 35'')$ are consistent and
conserved owing to the motion equations (29) and (35). Due to these
 properties the constraints (23) and $(24'-24'')$  can be considered
as the constraints for initial data  of  Eqs.(23) and Eqs.(24).
Generally covariant formulation of Eqs.(26) and  $(26'-26'')$ for the
second-order perturbative functions  $\chi^{M}(T,\sigma)$ should be
studied by analogy, but here we restrict ourselves by considering the
perturbative string dynamics in the first and second approximations.

As follows from the general covariant  formulation (35), the string
equations in the first approximation acquire  a simple form of a covariant
wave equation \begin{equation}\label{39} \left( {\cal
D}^{2}_{T}-\partial^{2}_{\sigma}\right)\psi^{M}=0 \end{equation} for the
class of symmetric spaces characterized by  the condition
\begin{equation}\label{40}
R_{MPQL}=\kappa (G_{MQ}G_{PL}-G_{ML}G_{PQ})
\end{equation}
Such a simplification is a consequence of the constraints $(24'')$ and
$(23')$, since
\begin{equation}\label{41}
R^{M}_{\ PQL}\varphi^{P}_{,T}\varphi^{Q}_{,T}\psi^{L}=
\kappa\left[ \varphi^{M}_{,T}\left(\psi_{N}\varphi^{N}_{,T}\right)-
\psi^{M}\left( \varphi_{N,T}\varphi^{N}_{,T}\right)\right]=0
\end{equation}

 The de Sitter space is an important example of the class of symmetric
spaces and will be studied below.

\section{Stability of perturbative string oscillations in the de Sitter
universe}

Here we consider the application of the above considered realization of
perturbative approach for the solution of the string
equations in the Friedmann-Robertson-Walker cosmological spaces. The F-R-W
metrics are characterized by the following quadratic form
\begin{equation}\label{42}
ds^{2}=\left(dx^{0}\right)^{2}-R^{2}(x^{0})\delta_{ik}dx^{i}dx^{k}
\end{equation}
The solution of the zero approximation equations and constraints $(23,
23')$  for the metric (42) is well-known and  has the form ( in notations
of [11]) \begin{equation}\label{43} \begin{eqalign}
T=T_{0}+\left(\nast^{0}\right)^{-1}
\int\nolimits_{\varphi^{0}(T_{0})}^{\varphi^{0}(T)}dt\, R(t), \\
\varphi^{i}(\varphi^{0})=\varphi^{i}(T_{0})+\nu^{i}
\int\nolimits_{\varphi^{0}(T_{0})}^{\varphi^{0
}(T)}dt\, R^{-1}(t),
\end{eqalign}
\end{equation}
where $\nu^{i}= \nast^{i}/\nast^{0}$  and  $ \nast^{0}(T_{0})$  are the
Cauchy initial data having the dimensionality  $L$. In terms of  these
initial data the constraint $(23')$ has the form
\begin{equation}\label{44} \nast^{M}\nast_{M}=0 \quad,\quad
\nu^{i}\nu^{i}=1 \end{equation} and  the tangent vectors  $
\varphi^{M}_{,T} $ are \begin{equation}\label{45}
\varphi^{0}_{,T}=\nast^{0}R^{-1}(\varphi^{0}) \quad,\quad
\varphi^{i}_{,T}=\nast^{i}R^{-2}(\varphi^{0})
\end{equation}

Now we may proceed  to the solution of Eqs.(35) and  the constraints
$(35'-35'')$. The substitution of the velocity components into the
constraints $(35'')$ appears to resolve them \begin{equation}\label{46}
\psi^{0}=R(\nu^{i}\psi^{i})
\end{equation}
To analyse the second constraint $(35')$ we may use  the
expressions for non-zero components of  $\Gamma^{M}_{PQ}(\varphi)$  for
the F-R-W metric (42)
\begin{equation}\label{47}
\begin{eqalign}
\Gamma^{0}_{ik}(\varphi)=\frac{1}{2}\delta_{ik}\partial_{0}R^{2}, \\
\Gamma^{i}_{0k}(\varphi)=\delta^{i}_{k}R^{-1}\partial_{0}R ,
\end{eqalign}
\end{equation}
where $ \partial_{0}\equiv\partial/{\partial\varphi^{0}}$. The covariant
derivatives (28) for the F-R-W metric are given by
\begin{equation}\label{48}
\begin{eqalign}
{\cal D}_{T}V^{0}=V^{0}_{,T}+\nast^{i}(R^{-1}\partial_{0}R)V^{i}, \\
{\cal D}_{T}V^{i}=V^{i}_{,T}+\nast^{0}(R^{-2}\partial_{0}R)V^{i}+
\nast^{i}(R^{-3}\partial_{0}R)V^{0}
\end{eqalign}
\end{equation}
After the substitution of (48) into the constraint $(35')$  we find that
the latter takes the form \begin{equation}\label{49} {\cal
D}_{T}\psi^{0}=R(\nu^{i}{\cal D}_{T}\psi^{i}) \end{equation} and is
identically satisfied by the solution (46).

Now let us leave the constraints and consider Eqs.(35). Using the
definitions (47-48) and the solutions (45) we find that Eqs.(35) are
transformed into the following ones
\begin{equation}\label{50}
\psi^{0}_{,TT}-\stackrel{\scriptstyle\prime\prime}{\psi}
\hspace{-3.5pt}\vphantom{\psi}^{0}+
2R^{-1}\partial_{0}R\nast^{i}\psi^{i}_{,T}+\nast^{i}\nast^{i}
\left[ \partial_{0}^{2}R+R^{-1}(\partial_{0}R)^{2} \right]
R^{-3}\psi^{0}=0,
 \end{equation}
 \begin{equation}\label{51}
\begin{eqalign}
\psi^{i}_{,TT}-\stackrel{\scriptstyle\prime\prime}{\psi}
\hspace{-3.5pt}\vphantom{\psi}^{i}+
2\nast^{0}R^{-2}\partial_{0}R\psi^{i}_{,T} +2\nast^{i}R^{-3}\partial_{0}R
\psi^{0}_{,T}+ \\
+2\nast_{0}\nast^{i}
\left[ \partial_{0}^{2}R-R^{-1}(\partial_{0}R)^{2} \right] R^{-4}\psi^{0}=0
\end{eqalign}
\end{equation}
After using the constraints (44) and (46) the equations (50-51) are
transformed into the separated ones for the time $\psi^{0}$ and space
$\psi^{i}$ components of $\psi^{M}$ \begin{equation}\label{52}
\psi^{0}_{,TT}-\stackrel{\scriptstyle\prime\prime}{\psi}
\hspace{-3.5pt}\vphantom{\psi}^{0}+ a\psi^{0}_{,T}+b\psi_{0}=0,
\end{equation}
\begin{equation}\label{53}
\psi^{i}_{,TT}-\stackrel{\scriptstyle\prime\prime}{\psi}
\hspace{-3.5pt}\vphantom{\psi}^{i}+
a_{ij}\psi^{j}_{,T}+b_{ij}\psi^{j}=0 ,
\end{equation}
where
\begin{equation}\label{54}
\begin{eqalign}
a=2\nast^{0}R^{-2}\partial_{0}R \quad,\quad
b=\left( \nast^{0}/R \right)^{2}\partial_{0}(R^{-1}\partial_{0}R) ,\\
a_{ij}=2\nast^{0}R^{-2}\partial_{0}R(\delta_{ij}+\nu_{i}\nu_{j}), \\
b_{ij}=2(\nast^{0})^{2}(R^{-3}\partial_{0}^{2}R)\nu_{i}\nu_{j}
\end{eqalign}
\end{equation}
Note that  Eqs.(50-51) have been obtained without use of the
constraint (46). Later it will be convenient to apply Eqs.(53) in some
other form similar to that of (51)
$$
\psi^{i}_{,TT}-\stackrel{\scriptstyle\prime\prime}{\psi}
\hspace{-3.5pt}\vphantom{\psi}^{i}+
a\psi^{i}_{,T}+ 2\nu^{i}\nast^{0}R^{-2}(R^{-1}\partial_{0}R\psi^{0})_{,T}=0
                                                                 \eqno(51')$$
Of course, Eqs.(52-53) are not independent in view of the usage of (46).
It can be verified that  Eqs.(53) are reduced to  Eq.(52) after their
multiplication by $\nu_{i}$, summing up and  use of the  constraint (46).
This result is a justification test of the general conclusion concerning
the consistency of the equations and constraints of the perturbative
scheme.

Now we are ready to consider the most interesting case of the de Sitter
space inflationary metrics with the conformal factor $R$ (42) equal to
\begin{equation}\label{55}
R=e^{H\varphi^{0}(T)}
\end{equation}
For this case the solution (43) acquires the form
\begin{equation}\label{56}
\begin{eqalign}
\varphi^{0}(T)=H^{-1}\mbox{ln } {[}N^{0}(T+\Lambda){]}, \\
\varphi^{i}(T)=q_{0}^{i}-\nu^{i}H^{-1}e^{-H\varphi^{0}(T)}=
q_{0}^{i}-\frac{\nu^{i}}{HN^{0}(T+\Lambda)} ,
\end{eqalign}
\end{equation}
where the dimensionless constants $ Hq_{0}^{i} $ and $ N^{0}$ equal to
\begin{equation}\label{57}
q_{0}^{i}=\varphi^i(T_{0})+\nu^{i}H^{-1}e^{-H\varphi^{0}(T_{0})} \quad,\quad
N^{0}=\nast^{0}H
\end{equation}
Eq.(56) shows that an asymptotic  scale for the worldsheet  time $T~$
$~(T\gg 1/\varepsilon)$, where  the considered perturbative scheme
correctly works, corresponds to the asymptotic scale in the cosmic time
$\varphi^{0}$. The substitution of $R$ corresponding to the solutions
(56) \begin{equation}\label{58} R=N^{0}(T+\Lambda) \end{equation} into
Eq.(54) gives $b=0$, and Eq.(52) is reduced to \begin{equation}\label{59}
\psi^{0}_{,TT}-\stackrel{\scriptstyle\prime\prime}{\psi}
\hspace{-3.5pt}\vphantom{\psi}^{0}+
2(T+\Lambda)^{-1}\psi^{0}_{,T}=0
\end{equation}
The same substitution transforms Eqs.(53) into the
equations
\begin{equation}\label{60}
\psi^{i}_{,TT}-\stackrel{\scriptstyle\prime\prime}{\psi}
\hspace{-3.5pt}\vphantom{\psi}^{i}+
2(T+\Lambda)^{-1}\psi^{i}_{,T}=
-2(N^{0})^{-1}\nu^{i}(T+\Lambda)^{-2}\psi^{0}_{,T}
\end{equation}
We will see that after shifting $\psi^{i}$ into $ \Theta^{i} $
\begin{equation}\label{61}
\Theta^{i}=\psi^{i}-\nu^{i}R^{-1}\psi^{0}=\psi^{i}-\nu^{i}
\left(N^{0}(T+\Lambda)\right)^{-1}\psi^{0}
\end{equation}
Eqs.(60) are transformed into homogenous wave equations
\begin{equation}\label{62}
\Theta^{i}_{,TT}-\stackrel{\scriptstyle\prime\prime}{\Theta}\hspace{-3.5pt}
\vphantom{\Theta}^{i}+
2(T+\Lambda)^{-1}\Theta^{i}_{,T}=0 ,
\end{equation}
which coincide with the Eq.(59) for  $\psi^{0}$. The proof is based
on an interesting property of the linear differential operator $\hat{L}$
in (59) and (60)
\begin{equation}\label{63}
\hat{L}=\partial_{T}^{2}-\partial_{\sigma}^{2}+2(T+\Lambda)^{-1}\partial_{T} ,
\end{equation}
which is expressed by a commutator relation
\begin{equation}\label{64}
\left[ \hat{L},(T+\Lambda)^{-1} \right]=-2(T+\Lambda)^{-2}\partial_{T}
\end{equation}
Following from Eq.(64)  is the relation
\begin{equation}\label{65}
\hat{L}\left((T+\Lambda)^{-1}\psi^{0}\right)=-2(T+\Lambda)^{-2}\psi^{0}_{,T},
\end{equation}
which proves the validity of Eqs.(62). As concerns the constraint (46),
it takes a simple form \begin{equation}\label{66} \nu^{i}\Theta^{i}=0
\end{equation}

Thus, we conclude that the perturbative string  dynamics in the first
approximation on $\varepsilon$ in the de Sitter space is described by
Eqs.(59,62) and the constraint (66).

To solve Eq.(62) take into account the periodicity condition with
respect to  $\sigma$ (22)
\begin{equation}\label{67}
 \Theta^{i}(T,0)=\Theta^{i}(T,2\pi) \quad,\quad
\psi^{0}(T,0)=\psi^{0}(T,2\pi)
\end{equation}
and expand $\Theta^{i}$ in a Fourier series
\begin{equation}\label{68}
\Theta^{i}=\sum_{n=-\infty}^{\infty}A^{i}_{n}(T)e^{in\sigma}
\end{equation}
After the substitution of the expansion (68) into the Eq.(62) the latter
transforms  into the Bessel equation for the Fourier coefficients
$a^{i}_{n}(T) $
\begin{equation}\label{69}
A^{i}_{n,TT}+\frac{2}{T+\Lambda}A^{i}_{n,T}+n^{2}A^{i}_{n}=0
\end{equation}
The general solution of Eqs.(69) for $n\neq0$ is [23]
\begin{equation}\label{70}
\begin{eqalign}
A^{i}_{n}(T)=(T+\Lambda)^{-1/2}Z_{-1/2}
\left(|n|(T+\Lambda)\right)= \\
=(T+\Lambda)^{-1/2}\left[ a^{i}_{n}J_{-1/2}\left(|n|(T+\Lambda)\right)+
b^{i}_{n}Y_{-1/2}\left(|n|(T+\Lambda)\right) \right] ,
\end{eqalign}
\end{equation}
where $J_{-1/2}(z) $ and $Y_{-1/2}(z)$ are the Bessel  functions of the
first and second  type respectively
\begin{equation}\label{71}
\begin{eqalign}
J_{-1/2}(z)=\left(\frac{2}{\pi z}\right)^{1/2}\cos z  ,\\
Y_{-1/2}(z)=J_{1/2}(z)=\left(\frac{2}{\pi z}\right)^{1/2}\sin z
\end{eqalign}
\end{equation}
and $a^{i}_{n}$  , $ b^{i}_{n}$ are the initial data.  For $ n=0$ the
general solution of Eqs.(69) is
\begin{equation}\label{72}
A^{i}_{0}=b^{i}_{0}+\frac{a^{i}_{0}}{T+\Lambda}
\end{equation}
Then the general solution of Eqs.(69) will be the following
$$
\Theta^{i}=
\frac{1}{T+\Lambda}\left\{ a^{i}_{0}+
\sum_{\vphantom{(} \atop {n=-\infty \atop n\not= 0}}^{\infty}
\left( \frac{2}{|n|\pi}\right)^{1/2}\!\!
\left[ a^{i}_{n}\cos |n|(T+\Lambda) +b^{i}_{n}\sin |n|(T+\Lambda)\right]
e^{in\sigma} \right\}
$$
\begin{equation}\label{73}
+b^{i}_{0}
\end{equation}
Introducing  new oscillator coefficients $\alpha^{i}_{n}$ and
$\beta^{i}_{n}$
\begin{equation}\label{74}
\begin{eqalign}
\alpha^{i}_{n}=(2\pi |n|)^{-1/2}(a^{i}_{n}+ib^{i}_{n})e^{-in\Lambda}, \\
\beta^{i}_{n}=(2\pi |n|)^{-1/2}(a^{i}_{n}-ib^{i}_{n})e^{in\Lambda}
\quad (n\not= 0), \\
\alpha^{i}_{0}=a^{i}_{0} \quad,\quad\alpha^{i}_{-n}=
\stackrel{*}{\alpha}\hspace{-3.5pt}\vphantom{\alpha}^{i}_{n} \quad,\quad
\beta^{i}_{-n}=
\stackrel{*}{\beta}\hspace{-3.5pt}\vphantom{\beta}^{i}_{n} \quad,\quad
\beta^{i}_{0}=0 ,
\end{eqalign}
\end{equation}
we present the solution (73) in the form of independent left and right
vawes running along  the closed string
\begin{equation}\label{75}
\begin{eqalign}
\Theta^{i}(T,\sigma)=\frac{1}{T+\Lambda}
\sum_{n=-\infty}^{\infty}\left[\alpha^{i}_{n}e^{in(\sigma-T)}+\beta^{i}_{n}
e^{in(\sigma+T)}\right]+b^{i}_{0}
\end{eqalign}
\end{equation}
The general solution of Eq.(59) has the same form
\begin{equation}\label{76}
\psi^{0}(T,\sigma)=\frac{1}{T+\Lambda}
\sum_{n=-\infty}^{\infty}\left[
\alpha^{0}_{n}e^{in(\sigma-T)}+\beta^{0}_{n}
e^{in(\sigma+T)}\right]+b^{0}_{0}
\end{equation}
The substitution of $\Theta^{i}$ (75) and $\psi^{0}$ (76) into the
representation (61) and the constraint (66) leads to the solution for
$\psi^{i}(T, \sigma)$
\begin{equation}\label{77}
\begin{eqalign}
\psi^{i}(T,\sigma)=\frac{1}{T+\Lambda}\sum_{n=-\infty}^{\infty}\left[
\alpha^{i}_{n}e^{in(\sigma-T)}+\beta^{i}_{n} e^{in(\sigma+T)}\right]+\\
+\frac{\nu^{i}}{N^{0}(T+\Lambda)^{2}}
\sum_{n=-\infty}^{\infty}\left[
\alpha^{0}_{n}e^{in(\sigma-T)}+\beta^{0}_{n}
e^{in(\sigma+T)}\right]+
\frac{\nu^{i}}{N^{0}(T+\Lambda)}b^{0}_{0}+b^{i}_{0}
\end{eqalign}
\end{equation}
together with the constraint for the
 oscillator coefficients  $\alpha^{i}_{n}$ and  $\beta^{i}_{n}$
$$
\nu^{i}\alpha^{i}_{n}=\nu^{i}\beta^{i}_{n}=\nu^{i}b^{i}_{0}=0   \eqno(78')
$$
Substituting the solution (55-56) and (76-77) into the expansion (20) we
find
\begin{equation}\label{78}
\begin{eqalign}
x^{0}(T,\sigma)=(H^{-1}\mbox{ln }N^{0}+\varepsilon b^{0}_{0})+\\
+\left\{H^{-1}\mbox{ln }(T+\Lambda)+\frac{\varepsilon}{T+\Lambda}
\sum_{n=-\infty}^{\infty}\left(
\alpha^{0}_{n}e^{in(\sigma-T)}+\beta^{0}_{n}
e^{in(\sigma+T)}\right)\right\} +O(\varepsilon^{2}) ,\\
x^{i}(T,\sigma)=(q^{i}_{0}+\varepsilon b^{i}_{0})+\\
+\frac{1}{N^{0}(T+\Lambda)}\left\{ \nu^{i}\biggl[-
H^{-1}+\varepsilon b^{0}_{0}+\frac{\varepsilon}{N^{0}(T+\Lambda)}
\sum_{n\not= 0}\left(
\alpha^{0}_{n}e^{in(\sigma-T)}+\beta^{0}_{n} e^{in(\sigma+T)}\right)\biggr]+
\right. \\
\left.
+\varepsilon\sum_{n\not= 0}\left(
\alpha^{i}_{n}e^{in(\sigma-T)}+\beta^{i}_{n} e^{in(\sigma+T)}\right)
\right\} +O(\varepsilon^{2})
\end{eqalign}
\end{equation}

 As follows from the representations (76) and (77), the perturbative
corrections  are connected with string oscillations in the directions
orthogonal and tangent to the geodesic  trajectory (56) of the zero
approximation. The amplitudes of these oscillations are asymptotically
small when  $T\gg1$ (or equivalently  $\tau\gg1/\varepsilon$) and the
amplitudes  of the longitudinal oscillations  are essentially smaller
than the amplitudes of the transversional oscillations. Therefore the
former oscillations may be neglected.

At the considered large scale $T$ the frequencies of perturbative
oscillations coincide with the Nambu-Goto frequencies. At the original
microscopic  scale $\tau$  these frequencies  are rescaled  by the
parameter $ \varepsilon$ and become  very small and equal to $
\varepsilon$, so we have
\begin{equation}\label{79}
\omega_{n}\Bigl|_{T\ {\rm scale}}=n \quad,\quad
\omega_{n}\Bigl|_{\tau\ {\rm scale}}=\varepsilon n
\end{equation}
As a consequence, all these frequencies  are stable at the considered large
scale  $T$ or equivalently when $ H\gg1/\sqrt{\alpha^{\prime}}$ (7).
    In the above discussion we  have already  noted that  the string
instabilities discovered in [5] was a consequence of the formula
\begin{equation}\label{80}
\omega_{n}=\sqrt{n^{2}-(\alpha'Hm)^{2}}
\end{equation}
for the oscillator frequencies, where the constant  $m$  is a
phenomenological mass parameter  associated with the mass of the particle
replacing the string in the zero approximation.  We have shown here that
in accordance with the variational principle based on the action
$S$ (1) the mass parameter $m$ must be equal to zero, and  this leads
to the disappearance of these instabilities. It seems that the condition
$m=0$ points out that such a parameter may be absent in string dynamics at
all. Indeed, it is known that a particle in de Sitter space has neither
definite mass nor definite spin, but has a definite eigenvalues of
two other Casimir operators. These eigenvalues are some combinations
of usual mass and angular momentum.

\section{Solution of the perturbative equations
\protect\\  in Friedmann-Robertson-Walker universes }

Here we shall study the perturbative  equations (52-54) in the F-R-W
universes  with a power parametrization of the scalar factor  $R$  (42)
\begin{equation}\label{81}
R=a(\varphi^{0})^{\alpha} ,
\end{equation}
where $a$ is a metric constant with the dimensionality $L^{-\alpha}$ and
$\alpha$ is an arbitrary parameter [3].

In this metric the solution (43) for  the cosmic time $\varphi^{0}(T)$
has the form
\begin{equation}\label{82}
\varphi^{0}(T)={\cal A}(T+\tilde{\Lambda})^{1/1+\alpha} \quad,\quad
(\alpha\not= -1) ,
\end{equation}
where the constant $A$ with the dimension $L$ and the dimensionless
constant $\tilde{\Lambda}$ are defined by the relations
\begin{equation}\label{83}
\begin{eqalign}
{\cal A}=\left( (1+\alpha)\nast^{0}a^{1/\alpha}
\right)^{1/1+\alpha}a^{-1/\alpha}, \\
\tilde{\Lambda}={\cal A}^{-(1+\alpha)}
\left( \varphi^{0}(T_{0})\right)^{\alpha+1} -T_{0}
\end{eqalign}
\end{equation}
For the space world coordinate $\varphi^{i}(T)$  the solution of Eq.(43)
is
\begin{equation}\label{84} \varphi^{i}\left(
\varphi^{0}(T))\right) =q^{i}_{0}+
\nu^{i}\left[(1-\alpha)a\right]^{-1}\left(\varphi^{0}(T)\right)^{1/1+\alpha}
\end{equation}
for ($\alpha\not=\pm 1 $). After the substitution of the representation (82)
into Eq.(84) we find
\begin{equation}\label{85}
\varphi^{i}(T)=q^{i}_{0}+
\nu^{i}B(T+\tilde{\Lambda})^{\frac{1-\alpha}{1+\alpha}}, \quad
(\alpha\not=\pm 1 ) ,
\end{equation}
where  the constants $q^{i}_{0}$ and $B$  are defined by
\begin{equation}\label{86}
\begin{eqalign}
q^{i}_{0}=\varphi^{i}(T_{0})-\frac{\nu^{i}}{a(1-\alpha)}
\left( \varphi^{0}(T_{0})\right)^{1-\alpha},
B=\frac{{\cal A}^{1-\alpha}}{a(1-\alpha)}
\end{eqalign}
\end{equation}
The special case $\alpha=\pm 1 $ may be easy studied separately.

Now let us to study the perturbative equations $(50-51')$. The substitution
of the solution (82-83) for $R$
\begin{equation}\label{87} R=a{\cal
A}^{\alpha}(T+\tilde{\Lambda})^{\alpha/1+\alpha} \qquad (\alpha\not= -1)
\end{equation}
into the relations defining the coefficients $ a,b, a_{ij}, b_{ij} $ (54)
gives the following expressions ($\alpha\not=\pm 1 $)
\begin{equation}\label{88}
\begin{eqalign}
a=\frac{2\alpha}{1+\alpha}(T+\tilde{\Lambda})^{-1}
\quad,\quad
a_{ij}=\frac{2\alpha}{1+\alpha}
(T+\tilde{\Lambda})^{-1}(\delta_{ij}+\nu_{i}\nu_{j}) \\
b=-\frac{\alpha}{(1+\alpha)^{2}}(T+\tilde{\Lambda})^{-2} \quad,\quad
b_{ij}=\frac{2\alpha(\alpha-1)}{(1+\alpha)^{2}}
(T+\tilde{\Lambda})^{-2}\nu_{i}\nu_{j}
\end{eqalign}
\end{equation}
Using (88) find that Eqs. (52,$51'$)  can be written  as
\begin{equation}\label{89}
\psi^{0}_{,TT}-\psi^{0}_{,\sigma\sigma}+
\frac{2\alpha}{1+\alpha}(T+\tilde{\Lambda})^{-1} \psi^{0}_{,T}-
\frac{\alpha}{(1+\alpha)^{2}}(T+\tilde{\Lambda})^{-2} \psi^{0}=0,
\end{equation}
\begin{equation}\label{90}
\begin{eqalign}
\psi^{i}_{,TT}-\psi^{i}_{,\sigma\sigma}+
\frac{2\alpha}{1+\alpha}(T+\tilde{\Lambda})^{-1} \psi^{i}_{,T}+\\
+\nu^{i}(T+\tilde{\Lambda})^{-\frac{1+2\alpha}{1+\alpha}}r\psi^{0}_{,T}-
\nu^{i}(T+\tilde{\Lambda})^{-\frac{2+3\alpha}{1+\alpha}}s\psi^{0}=0  ,
\end{eqalign}
\end{equation}
where  the constant coefficients $r$ and $s$ are defined as
\begin{equation}\label{91}
\begin{eqalign}
r=2\frac{\alpha}{(1+\alpha)^{\frac{1+2\alpha}{1+\alpha}}}
\left(a\cdot (\nast^{0})^{\alpha}\right)^{-1/1+\alpha},\\
s=2\frac{\alpha}{(1+\alpha)^{\frac{2+3\alpha}{1+\alpha}}}
\left(a\cdot (\nast^{0})^{\alpha}\right)^{-1/1+\alpha}=r/1+\alpha
\end{eqalign}
\end{equation}
The coefficients $r$ and $s$ are  dimensionless because they
include the dimensional constants $a$ and $\nast^{0} $ only in the
dimensionless combination  $ [a\cdot (\nast^{0})^{\alpha}]$. Introducing
a dimensionless constant $\kappa$ $$ \kappa=\left(
(1+\alpha)a^{1/\alpha}\nast^{0} \right)^{\frac{\alpha}{\alpha+1}} $$ we
can present the constraint (46) in the form
\begin{equation}\label{92}
\psi^{0}=\kappa
(T+\tilde{\Lambda})^{\frac{\alpha}{\alpha+1}}\left( \nu^{i}\psi^{i}\right)
\end{equation}
This constraint can be omitted now, because it was used for obtaining
Eqs.(90). So if we substitute the constraint (92) in Eq.(89) then the
latter transforms into a linear combination of Eqs.(90).  In the limiting
case when $\alpha^{\prime}\rightarrow\infty~ $ Eqs.(89-90) are reduced to
the de Sitter equations (59) and (69), because the coefficients $r$  and
$s$ (91) go to zero. These equations belong to the same  class of the
Bessel-like equations.

To solve Eq.(89) expand $\psi^{0}(T, \sigma) $ in a Fourier series
\begin{equation}\label{93}
\psi^{0}(T,\sigma)=
\tilde{\cal A}\vphantom{\cal A}_{0}^{0}(T)+
\sum_{n\not=0}\tilde{\cal A}\vphantom{\cal A}_{n}^{0}(T)e^{in\sigma} ,
\end{equation}
substitute the expansion (93) into  Eq.(89) and get the equation  for
 $ \tilde{\cal A}\vphantom{\cal A}_{n}^{0}(T) $
\begin{equation}\label{94}
\tilde{\cal A}\vphantom{\cal A}_{n,TT}^{0}+
\frac{2\alpha}{1+\alpha}(T+\tilde{\Lambda})^{-1}\tilde{\cal A}^{0}_{n,T}+
\left[ n^{2}-\frac{\alpha}{(1+\alpha)^{2}}
(T+\tilde{\Lambda})^{-2}
\right]\tilde{\cal A}^{0}_{n}=0
\end{equation}

The general solution  of Eq.(94) for the case  $n\neq 0~$ has the form
(see [23])
\begin{equation}\label{95}
\begin{eqalign}
\tilde{\cal A}\vphantom{\cal A}_{n}^{0}(T)=
(T+\tilde{\Lambda})^{\frac{1-\alpha}{2(1+\alpha)}}Z_{-1/2}
\left( |n|(T+\tilde{\Lambda})\right)= \\
=(T+\tilde{\Lambda})^{\frac{1-\alpha}{2(1+\alpha)}}
\left[
\tilde{a}\vphantom{a}^{0}_{n}J_{-1/2}\left(|n|(T+\tilde{\Lambda})\right)+
\tilde{b}\vphantom{b}^{0}_{n}Y_{-1/2}\left(|n|(T+\tilde{\Lambda})\right)
\right]
\end{eqalign}
\end{equation}
Respectively the general solution for  $\tilde{\cal
A}\vphantom{\cal A}_{0}^{0}(T)$ corresponding the case $n=0 $ is
\begin{equation}\label{96}
 \tilde{\cal A}\vphantom{\cal A}_{0}^{0}(T)=
\tilde{a}\vphantom{a}_{0}^{0}
(T+\tilde{\Lambda})^{\frac{-\alpha}{1+\alpha}}+
\tilde{b}\vphantom{b}_{0}^{0}
(T+\tilde{\Lambda})^{\frac{1}{1+\alpha}}
\end{equation}

The substitution of the solutions (95-96) into (93) gives the general
solution of Eq.(89) \begin{equation}\label{97} \psi^{0}(T,\sigma)=
(T+\tilde{\Lambda})^{\frac{-\alpha}{1+\alpha}}
\sum_{n=-\infty}^{\infty}\left[ \tilde{\alpha}
\vphantom{\alpha}^{0}_{n}e^{in(\sigma-T)}+
\tilde{\beta}\vphantom{\beta}^{0}_{n}e^{in(\sigma+T)}\right] +
\tilde{b}\vphantom{b}_{0}^{0}
(T+\tilde{\Lambda})^{\frac{1}{1+\alpha}}
\end{equation}
In the limiting case when $\alpha^{\prime}\rightarrow\infty $ the solution
(97) reduces to the solution (76). The general solution (20) for the
cosmic time coordinate $x^{0}(T,\sigma) $ acquires the form

\begin{equation}\label{98}
\begin{eqalign}
x^{0}(T,\sigma)=\left[{\cal A}+ \varepsilon
\tilde{b}\vphantom{b}_{0}^{0}\right](T+\tilde{\Lambda})^{\frac{1}{1+\alpha}}+
\\
+\varepsilon(T+\tilde{\Lambda})^{-\frac{\alpha}{1+\alpha}}
\sum_{n=-\infty}^{\infty}\left[ \tilde{\alpha}
\vphantom{\alpha}^{0}_{n}e^{in(\sigma-T)}+
\tilde{\beta}\vphantom{\beta}^{0}_{n}e^{in(\sigma+T)}\right] +
O(\varepsilon^{2})
\end{eqalign}
\end{equation}

Having the solution (97) for $~\psi^{0}(T)$ we shall seek for the  general
solution of Eqs.(90) in the form of the Foutier series expansion
\begin{equation}\label{99}
\psi^{i}(T,\sigma)=
\tilde{\cal A}\vphantom{\cal A}_{0}^{i}(T)+
\sum_{n\not=0}\tilde{\cal A}\vphantom{\cal A}_{n}^{i}(T)e^{in\sigma} ,
\end{equation}
Then the substitution of the expansions (99) and (93) into Eqs.(90)
will give the equations for $\tilde{\cal A}\vphantom{\cal
A}_{0}^{i}(T) $ and $ \tilde{\cal A}\vphantom{\cal A}_{n}^{i}(T)$. We find
that the equation for the zero mode $\tilde{\cal A}\vphantom{\cal
A}_{i}^{0}(T)$ turns out to be the following
\begin{equation}\label{100}
\begin{eqalign}
\tilde{\cal A}\vphantom{\cal A}_{0,TT}^{i}+
\frac{2\alpha}{1+\alpha}(T+\tilde{\Lambda})^{-1} {\cal A}^{i}_{0,T}=
\nu^{i}r\tilde{a}\vphantom{a}_{0}^{0}(T+\tilde{\Lambda})
^{\frac{-2(1+\alpha)}{1+\alpha}}
\end{eqalign}
\end{equation}
After rewriting Eq.(100) in the form
\begin{equation}\label{101}
\begin{eqalign}
(T+\tilde{\Lambda})^{\frac{-2\alpha}{1+\alpha}}
 \left[ (T+\tilde{\Lambda})^{\frac{2\alpha}{1+\alpha}}
  \tilde{\cal A}\vphantom{\cal
  A}_{0,T}^{i}\right]_{,T}=\nu^{i}r\tilde{a}\vphantom{a}_{0} ^{0}(T+\tilde{\Lambda})
^{\frac{-2(1+\alpha)}{1+\alpha}}
\end{eqalign}
\end{equation}
it is easily integrated, and its general solution is
\begin{equation}\label{102}
\begin{eqalign}
\tilde{\cal A}\vphantom{\cal
A}_{0}^{i}={\frac{1+\alpha}{2\alpha}}
\nu^{i}r\tilde{a}\vphantom{a}_{0} ^ {0}(T+\tilde{\Lambda})
^{\frac{-2\alpha}{1+\alpha}} + {\frac{1+\alpha}{1-\alpha}}
\tilde{\cal C}\vphantom{\cal
C}_{01}^{i}(T+\tilde{\Lambda})^{\frac{1-\alpha}{1+\alpha}} +\tilde{\cal
C}\vphantom{\cal C}_{02}^{i}
 \end{eqalign}
 \end{equation}
Similarly one can derive the equation for the n-th mode
$\tilde{\cal A}\vphantom{\cal A}_{n}^{i}(T)$
\begin{equation}\label{103}
\begin{eqalign}
\tilde{\cal A}\vphantom{\cal A}_{n,TT}^{i}+
\frac{2\alpha}{1+\alpha}(T+\tilde{\Lambda})^{-1} {\cal A}^{i}_{n,T}+n^{2}
\tilde{\cal A}\vphantom{\cal A}_{n}^{i}=\nu^{i}rF_{n}  ,
\end{eqalign}
\end{equation}
where $F_{n}$ is defined  as
\begin{equation}\label{104}
\begin{eqalign}
F_{n}\equiv-(T+\tilde{\Lambda})^{-\frac{1+2\alpha}{1+\alpha}}\left[
\tilde{\cal A}\vphantom{\cal A}_{n,T}^{0}-\frac{\alpha}{1+\alpha}
(T+\tilde{\Lambda})^{-1}\tilde{\cal A}\vphantom{\cal A}_{n}^{0}\right]
\end{eqalign}
\end{equation}

The substitution of $\tilde{\cal A}\vphantom{\cal A}_{n}^{0}$ (95)
into Eq.(104) allows to present $F_{n}$ as
\begin{equation}\label{105}
\begin{eqalign}
F_{n}=\sqrt{\frac{2}{\pi|n|}}(T+\tilde{\Lambda})^{-\frac{5+9\alpha}
{2(1+\alpha)}} \lbrace \tilde{f}
\vphantom{f}^{0}_{1n}
\left[ \cos
|n|(T+\tilde{\Lambda}) + |n|(T+\tilde{\Lambda})\sin|n|(T+\tilde{\Lambda})
\right] +\\ + \tilde{f} \vphantom {f}^{0}_{2n} \left[
\sin|n|(T+\tilde{\Lambda})
-|n|(T+\tilde{\Lambda})\cos|n|(T+\tilde{\Lambda}) \right] \rbrace
\end{eqalign}
\end{equation}
The general solution of Eq.(103) is a sum of the general solution of the
gomogenious equation
\begin{equation}\label{106}
\begin{eqalign}
\tilde{\cal B}\vphantom{\cal B}_{n,TT}^{i}+
\frac{2\alpha}{1+\alpha}(T+\tilde{\Lambda})^{-1} {\cal B}^{i}_{n,T}+n^{2}
\tilde{\cal B}\vphantom{\cal B}_{n}^{i}=0
\end{eqalign}
\end{equation}
and a particular solution of Eq.(103). The general solution of Eq.(106) is
given by the expression
\begin{equation}\label{107}
\begin{eqalign}
\tilde{\cal B}\vphantom{\cal B}_{n}^{i}=
(T+\tilde{\Lambda})^{\frac{1-\alpha}{2(1+\alpha)}}
Z_{\frac{1-\alpha}{2(1+\alpha)}}
\left( |n|(T+\tilde{\Lambda})\right)= \\
=(T+\tilde{\Lambda})^{\frac{1-\alpha}{2(1+\alpha)}}
\left[ \tilde{b}\vphantom{b}^{i}_{1n}J_{\frac{1-\alpha}{2(1+\alpha)}}
\left(|n|(T+\tilde{\Lambda})\right)+
\tilde{b}\vphantom{b}^{i}_{2n}Y_{\frac{1-\alpha}{2(1+\alpha)}}
\left(|n|(T+\tilde{\Lambda})\right) \right] ,
\end{eqalign}
\end{equation}
where $ J_{\frac{1-\alpha}{2(1+\alpha)}}(z)$  and
$Y_{\frac{1-\alpha}{2(1+\alpha)}}(z)$ are the Bessel functions of the
first and second order respectively.

The general solution of Eqs.(103) is presented as a sum of the solution
$\tilde{\cal B}\vphantom{\cal B}_{n}^{i}$ (107) and a particular  solution
$\tilde{\cal H}\vphantom{\cal H}_{n}^{i}$
\begin{equation}\label{108}
\begin{eqalign}
\tilde{\cal A}\vphantom{\cal A}_{n}^{i}= \gamma_{n}
\tilde{\cal B}\vphantom{\cal B}_{n}^{i}+
\tilde{\cal H}\vphantom{\cal H}_{n}^{i}
\end{eqalign}
\end{equation}
Eq.(103) with the right-hand side given by the expression (105) belongs to
the set of exactly integrable inhomogenious  Bessel-like equations.
Therefore the particular solution $\tilde{\cal H}\vphantom{\cal H}_{n}^{i}
$ (108), depending on the parameter $\alpha$, belongs to the set of
one-parameter solutions discussed in [23]. That is why we do not dwell
on the discussion of these particular solutions. Instead we
shall show another way for the solution of Eqs.(90).

To this end notice that
Eqs.(90) as well as $ Eqs.(51') $ preceding them,
mix the space component $\psi^{i}$ with the time component $\psi^{0}$ and
its T-derivative. On the other hand, Eqs.(53) are equivalent to $
Eqs.(51') $ and contain only the space component $\psi^{i}$.
Thus, we can solve Eqs.(53) independently on $\psi^{0}$ and then use
the constraint (46) (or (98)) for establishing a connection between the
integration constants  contained in the solutions for $\psi^{i}$  and
$\psi^{0}$.

To illustrate this possibility we consider the  special case of
initial data when the velocity $\nu^{i}$ has only one non-zero
 component $\nu^{z}$, i.e.
\begin{equation}\label{109}
\begin{eqalign}
\nu^{i}\equiv(\nu^{x},\nu^{y},\nu^{z})\equiv(\nu^{t},\nu^{z})=(0, 0, 1)
\end{eqalign}
\end{equation}
The general case of arbitrary initial data  for $\nu^{i}$  reduces to
the case (109) after the fixation of the coordinate frame in
the F-R-W space-time. Such a choice of the general covariant gauge is a
correct operation in view of the general covariance of the perturbative
scheme studied here. In the gauge (109) Eqs.(53) transform to the
homogenious Bessel-like equations \begin{equation}\label{110}
\psi^{t}_{,TT}-\psi^{t}_{,\sigma\sigma}+
\frac{2\alpha}{1+\alpha}(T+\tilde{\Lambda})^{-1} \psi^{t}_{,T}=0,
\end{equation}
\begin{equation}\label{111}
\psi^{z}_{,TT}-\psi^{z}_{,\sigma\sigma}+
\frac{4\alpha}{1+\alpha}(T+\tilde{\Lambda})^{-1} \psi^{z}_{,T}-
\frac{2\alpha(1-\alpha}{(1+\alpha)^{2}}(T+\tilde{\Lambda})^{-2}
\psi^{z}=0,
\end{equation}
After the substitution of the $ \psi^{t}$ Fourier expansion
\begin{equation}\label{112}
\psi^{t}(T,\sigma)=
\tilde{\cal A}\vphantom{\cal A}_{0}^{t}(T)+
\sum_{n\not=0}\tilde{\cal A}\vphantom{\cal A}_{n}^{t}(T)e^{in\sigma} ,
\end{equation}
into Eqs.(110) we get the solutions (102) and (107) for the Fourier
components $\tilde{\cal A}\vphantom{\cal A}_{0}^{t}(T)$ and
$\tilde{\cal A}\vphantom{\cal A}_{n}^{t}(T)$
\begin{equation}\label{113}
\tilde{\cal A}\vphantom{\cal A}_{0}^{t}={\frac{1+\alpha}{1-\alpha}}
\tilde{\cal C}\vphantom{\cal
C}_{01}^{t}(T+\tilde{\Lambda})^{\frac{1-\alpha}{1+\alpha}}+ \tilde{\cal
C}\vphantom{\cal C}_{02}^{t}  ,
\end{equation}
\begin{equation}\label{114}
\begin{eqalign}
\tilde{\cal A}\vphantom{\cal A}_{n}^{t}=
(T+\tilde{\Lambda})^{\frac{1-\alpha}{2(1+\alpha)}}
Z_{\frac{1-\alpha}{2(1+\alpha)}}
\left( |n|(T+\tilde{\Lambda})\right)= \\
=(T+\tilde{\Lambda})^{\frac{1-\alpha}{2(1+\alpha)}}
\left[ \tilde{b}\vphantom{a}^{t}_{1n}J_{\frac{1-\alpha}{2(1+\alpha)}}
\left(|n|(T+\tilde{\Lambda})\right)+
\tilde{b}\vphantom{b}^{t}_{2n}Y_{\frac{1-\alpha}{2(1+\alpha)}}
\left(|n|(T+\tilde{\Lambda})\right) \right] ,
\end{eqalign}
\end{equation}
The substitution of the Fourier expansion of $\psi^{z}$
\begin{equation}\label{115}
\psi^{z}(T,\sigma)=
\sum_{n}\tilde{\cal A}\vphantom{\cal A}_{n}^{z}(T)e^{in\sigma} ,
\end{equation}
into Eq.(111) transforms the latter into the equation
\begin{equation}\label{116}
\tilde{\cal A}\vphantom{\cal A}_{n,TT}^{z}+
\frac{4\alpha}{1+\alpha}(T+\tilde{\Lambda})^{-1}\tilde{\cal A}^{z}_{n,T}+
\left[ n^{2}-\frac{2\alpha(1-\alpha)}{(1+\alpha)^{2}}
(T+\tilde{\Lambda})^{-2}
\right]\tilde{\cal A}^{z}_{n}=0 ,
\end{equation}
which is similar to Eq.(94) for $\psi^{0}$. The general solution  of
Eq.(116) is
\begin{equation}\label{117}
 \tilde{\cal A}\vphantom{\cal A}_{0}^{z}(T)=
\tilde{a}\vphantom{a}_{0}^{z}
(T+\tilde{\Lambda})^{\frac{-2\alpha}{1+\alpha}}+
\tilde{b}\vphantom{b}_{0}^{z}
(T+\tilde{\Lambda})^{\frac{1-\alpha}{1+\alpha}}
\end{equation}
for the zero mode of the expansion (115) and
\begin{equation}\label{118}
\begin{eqalign}
\tilde{\cal A}\vphantom{\cal A}_{n}^{z}(T)=
(T+\tilde{\Lambda})^{\frac{1-3\alpha}{2(1+\alpha)}}Z_{-1/2}
\left( |n|(T+\tilde{\Lambda})\right)= \\
=(T+\tilde{\Lambda})^{\frac{1-3\alpha}{2(1+\alpha)}}
\left[
\tilde{a}\vphantom{a}^{z}_{n}J_{-1/2}\left(|n|(T+\tilde{\Lambda})\right)+
\tilde{b}\vphantom{b}^{z}_{n}Y_{-1/2}\left(|n|(T+\tilde{\Lambda})\right)
\right]
\end{eqalign}
\end{equation}
for the oscillatory modes.

Now let us return to the constraint (92) and substitute  the solutions
(113-114), (117-118) and (95-96) in this constraint. We find that
the constraint (92)   will be identically satisfied , if the integration
constants are connected by the relations
\begin{equation}\label{119}
\begin{eqalign}
\tilde{a}\vphantom{a}^{0}_{n}=\kappa\tilde{a}\vphantom{a}^{z}_{n}, \quad
\tilde{b}\vphantom{b}^{0}_{n}=\kappa\tilde{b}\vphantom{b}^{z}_{n}
\end{eqalign}
\end{equation}
for all $n$.
Note that the constraint (92) does not restrict the integration constants
contained  in the solutions  (113-114)  which describe the string
oscillations orthogonal to the initial velocity $\nu^{i}$ (109). Taking
into account the relations (119)  we find that the solution for $\psi^{z}$
can be presented in the form similar to the one given by (97).
\begin{equation}\label{120}
\psi^{z}(T,\sigma)=\kappa
(T+\tilde{\Lambda})^{{\frac{1-\alpha}{1+\alpha}}-1}
\sum_{n=-\infty}^{\infty}\left[ \tilde{\alpha}
\vphantom{\alpha}^{0}_{n}e^{in(\sigma-T)}+
\tilde{\beta}\vphantom{\beta}^{0}_{n}e^{in(\sigma+T)}\right] +
\kappa\tilde{b}\vphantom{b}_{0}^{0}
(T+\tilde{\Lambda})^{\frac{1-\alpha}{1+\alpha}}
\end{equation}
Substituting the solutions (85) and (120) into the perturbative expansions
(20) we obtain the following  solution for the world string  coordinate $
x^{z}$
\begin{equation}\label{121}
\begin{eqalign}
x^{z}(T,\sigma)=q^{z}_{0}+ \left[B+ \varepsilon\kappa
\tilde{b}\vphantom{b}_{0}^{0}\right](T+\tilde{\Lambda})^{\frac{1-\alpha}
{1+\alpha}}+
\\
+\varepsilon\kappa(T+\tilde{\Lambda})^{\frac{1-\alpha}{1+\alpha}-1}
\sum_{n=-\infty}^{\infty}\left[ \tilde{\alpha}
\vphantom{\alpha}^{0}_{n}e^{in(\sigma-T)}+
\tilde{\beta}\vphantom{\beta}^{0}_{n}e^{in(\sigma+T)}\right] +
O(\varepsilon^{2})
\end{eqalign}
\end{equation}
As follows from the representation  (121), the zero mode of $\psi^{z}$
gives a correction to the translational movement whereas the oscillations
of $z$ give rise to an additional oscillatory  movement of the string
coordinate $x^{z}$. Moreover in the asymptotic regime,
when $T\gg1$ (or equivalently  $\tau\gg1/\varepsilon$), the amplitude of
the oscillation are smaller than the corresponding translation . This
behaviour of $x^{z}$ is in agreement with the above-mentioned qualitive
picture of the perturbative string dynamics in curved space-time.

Taking into account a weak tension also leads to the appearance of the
translations
(113) and the oscillations (114) in the transverse directions  to the
$\nu^{i}$ velocity (109). In the zero approximation the string motion in
these transverse directions  was absent.  A general of the
amplitudes of the string oscillations in $x, y$ and $z$ directions is thus
asymptotic drop when the parameter $\alpha$ lies in the region
$\alpha>0 $ or $\alpha<-1 $.

At this point we stop our general discussion illustrating the
applicability of the proposed realization for the perturbative approach.
More detailed analysis of the perturbative
string motions depends on the values of $\alpha$ and the initial data
contained into the presented solutions. We shall return to this analysis
in another paper.

\section{Conclusion}
We discuss here the problem of approximate solution of the string
equations in curved space-time. A suitable representation for the
string action with covariantly separated kinetic and potential terms is
applied for this goal. Using the existence of a dimensional parameter in
the metric of curved space a dimensionless parameter depending on the
string tension is built. It is shown that the potential term in the string
action can be treated as a perturbation for the case of smallness of this
dimensionless parameter. At the same time this small parameter is
appearing in the constraints and Euler-Lagrange variational equations and
they can be reformulated into the chain of perturbative linear
equations.

Established is the fact that the perturbative string equations for the
de Sitter and the Friedmann-Robertson-Walker universes are reduced to the
linear system of the exactly solvable modified Bessel equations. Moreover,
the corresponding string constraints are transformed to the simple linear
conditions for the Fourier coefficients in the expansions of the
perturbative solutions. The proposed approximation selfconsistently
describes the string dynamics on the scale of large values for the
world-sheet time in the fixed gauge. The asymptotic non-trivial string
motion has the character of damped oscillations with the amplitudes
falling as a power of the slow worldsheet time. An interesting peculiarity
of this perturbative description is the asymptotic stability of the string
dynamics in the de Sitter space for a large Hubble constant.

\section{Acknowledgement}

We would like to thank M.P. Dabrowski, A. Larsen, U. Lindstrom,
A. Nicolaidis, I.D. Novikov,  N. Sanchez, A.A. Vodyanitskii and M. Zabzine 
for useful discussions.  This work is supported in part by INTAS Grant N 
93-127-ext and in part by SFFI Grants of Ukraine N $\Phi$4/1751 and N 
$\Phi$5/1794.


\end{document}